\documentclass[aps,prd,showpacs,twocolumn,floatfix,superscriptaddress]{revtex4}
\pdfoutput=1

\usepackage{amssymb,amsfonts}
\usepackage{bm} 
\usepackage{graphicx}

\newcommand{\salto}[1]{\left[\,#1\,\right]^{}_{p}}
\newcommand{\be}{\begin{equation}}
\newcommand{\ee}{\end{equation}}
\newcommand{\bea}{\begin{eqnarray}}
\newcommand{\eea}{\end{eqnarray}}
\newcommand{\nn}{\nonumber}

\newcommand{\Phifield}{\varphi}
\newcommand{\Pifield}{\phi}
\newcommand{\rstar}{{r^\ast}}
\newcommand{\rstarp}{{r^\ast_{p}}}
\newcommand{\Smooth}{{\mbox{\tiny S}}}

\begin{document}

\title{Are Time-Domain Self-Force Calculations Contaminated by Jost Solutions?}

\newcommand*{\IEEC}{Institut de Ci\`encies de l'Espai (CSIC-IEEC), Facultat
de Ci\`encies, Campus UAB, Torre C5 parells, 08193 Bellaterra, Spain}

\newcommand*{\AEI}{Max Planck Institut f\"ur Gravitationsphysik, 
Albert-Einstein-Institut, 14476 Potsdam, Germany} 

\newcommand*{\MEU}{Laboratoire Univers et Th\'eories (LUTH), Observatoire de Paris, CNRS, 
Universit\'e Paris Diderot, 92190 Meudon, France} 

\author{Jos\'e Luis Jaramillo}
\email{Jose-Luis.Jaramillo@aei.mpg.de} 
\affiliation{\AEI}
\affiliation{\MEU}
\author{Carlos F. Sopuerta}
\email{sopuerta@ieec.uab.es} 
\affiliation{\IEEC}
\author{Priscilla Canizares}
\email{pcm@ieec.uab.es}
\affiliation{\IEEC}

\date{\today}

\begin{abstract}
The calculation of the self force in the modeling of the gravitational-wave emission from
extreme-mass-ratio binaries is a challenging task.  Here we address the
question of the possible emergence of a persistent spurious solution in
time-domain schemes, referred to as a {\em Jost junk solution} in the literature,
that may contaminate self force calculations.  Previous studies suggested that
Jost solutions are due to the use of zero initial data, which is inconsistent
with the singular sources associated with the small object, described as a point mass.
However, in this work we show that the specific origin is an inconsistency in the translation
of the singular sources into jump conditions.  More importantly, we
identify the correct implementation of the sources at late times as the sufficient
condition guaranteeing the absence of Jost junk solutions.

\end{abstract}

\pacs{04.25.dg, 04.30.Db, 95.30.Sf}

\maketitle

Extreme-mass-ratio inspirals are one of the most important sources of gravitational
radiation for the future space-based observatory Laser Interferometer Space Antenna 
(LISA). They consist of a stellar compact object inspiralling into a 
massive black hole (MBH; with mass in the range $M = 10^{4}-10^{7}M_{\odot}$).  
They are long lasting sources that in the last year before plunge can spend of the 
order of $10^{5}$ cycles inside the LISA frequency band~\cite{Finn:2000sy}.  To 
extract these signals from the future LISA data stream we require very precise 
theoretical waveform templates, as the signal-to-noise ratio accumulates slowly with 
time~\cite{Gair:2004iv}.  To achieve such precision we need a very accurate 
description of the slow inspiral, which can be seen as due to the action of a local 
force, the {\em self force}.   The computation of the self force is a challenge 
since we need to regularize the gravitational perturbations created by the stellar 
compact object, which is modeled as a point mass.  Nevertheless, during the last 
decade there has been significant progress in the self-force program 
(see~\cite{Poisson:2004lr,Barack:2009ux} for reviews) and different methods to 
compute it have been developed.  Of particular relevance is the computation
of the gravitational self force in the case of a nonrotating MBH, both for
circular and eccentric orbits~\cite{Barack:2007tm,Barack:2010tm}, using time-domain 
methods.  

The aim of this letter is to clarify an issue related to time-domain 
self-force calculations: The appearance of the so-called {\em Jost junk}
solutions discussed recently in~\cite{Field:2010xn}, which can contaminate
time-domain calculations of the self force, limiting their accuracy.
Here, following~\cite{Field:2010xn}, we show that such solutions are not a generic feature 
of time-domain schemes, but rather a result of the use of trivial zero initial data 
combined with a particular implementation of the singular sources associated with the
point mass.  We then identify the specific ingredients responsible for the eventual
appearance of the spurious solution and point out a straightforward solution,
eliminating in this way concerns about the generation of this kind of
persistent junk contamination when calculating the self force.

The self force is determined by the gravitational perturbations created by the point
mass as it orbits the MBH, what we call the retarded field, whose computation is
the main numerical task in time-domain self-force schemes.  The retarded field
is singular at the location of the point mass and needs to be regularized by
using analytic expressions for the singular field (see, e.g.~\cite{Barack:2001gx}).  
The retarded field is described by a system of coupled wave-type equations that derive 
from the perturbed Einstein's equations, which means that their structure depends on the
gauge adopted.  Not all the gauges are suitable for self-force computations.

The principal part of the wave-type equations that govern the retarded field is common
to most gauges.  For nonrotating MBHs, each spherical harmonic component obeys decoupled
$1+1$ wave equations whose structure is captured by the following model equation:
\bea
&&\left[-\partial_t^2 + \partial_\rstar^2 - V(r)\right] \Psi(t,r) = S(t,r) \nn \\
&& = f(r) \left[\, G(t,r)\,\delta(r-r^{}_p(t)) 
+ F(t,r)\,\delta'(r-r^{}_p(t)) \,\right]\,,  \label{e:master_equation}
\eea
where $f(r)= 1-2M/r$, $M$ is the MBH mass, $\rstar$ is the {\em tortoise} coordinate 
$\rstar = r+2M\mathrm{ln}[r/(2M) -1]$,  and $r_p(t)$ is the particle's radial motion.
This model equation consists of a $1+1$ wave operator $-\partial_t^2 + \partial_\rstar^2$,
a potential term, $V(r)\Psi$, and a singular source term $S(t,r)$.
Equation~(\ref{e:master_equation}) has the form of the master equations
that rule perturbations of different fields in Schwarzschild spacetime.  They are tied to
a particular set of gauges, but in other gauges the principal structure of the equations
that govern the dynamics of the perturbations is similar.  Therefore, we will use the model
Eq.~(\ref{e:master_equation}) in order to analyze the appearance of the Jost junk solution,
as done in~\cite{Field:2010xn}.  Without loss of generality, we deal with the case of
circular orbits ($r_p=\mathrm{const})$.

Time-domain schemes address the resolution of Eq.~(\ref{e:master_equation}) 
as an initial-boundary value problem, by prescribing initial conditions
on a time slice together with appropriate boundary conditions along the 
evolution. 
Appropriate initial data (ID) at a finite time are not known for 
Eq.~(\ref{e:master_equation}). A common practice consists of setting 
{\em trivial} ID, namely 
\be
\Psi|^{}_{t=t_o} = (\partial_t \Psi)|^{}_{t=t_o} = 0 \,. \label{e:trivial_data1}
\ee
This ID is inconsistent with the singular structure of the source 
and corresponds to a solution that is not continuous in time at $t=t_o$.
As a consequence, when  solving (numerically)
Eq.~(\ref{e:master_equation}) an initial burst of {\em junk radiation} is produced 
and one must wait until it has been radiated away in order to compute the self force. 
This strategy relies on the assumption that {\em junk radiation} is actually radiated away.
In this sense, Ref.~\cite{Field:2010xn} addresses the question of whether the use 
of trivial ID can give rise to spurious solutions persistent in time.  To answer this question
a double complementary numerical and analytical approach 
is adopted in~\cite{Field:2010xn} that we reanalyze below.

{\em (a) Numerical approach.}
The impact of inconsistent ID is assessed by constructing,
first, a solution to Eq. (\ref{e:master_equation}) with trivial ID, 
referred to as $\Psi_{\mathrm{Impulsive}}$. Second, the sources in 
Eq.~(\ref{e:master_equation}) are modified to make them compatible 
with trivial ID. This is achieved by a smooth switch on in time of the sources
\be
F_\tau^{\Smooth}(t,r)\equiv \alpha(t, \tau) F(t,r) \,,  \;
G_\tau^{\Smooth}(t,r)\equiv \alpha(t, \tau) G(t,r) \,,
\label{e:time_switch}
\ee
where $\alpha(t, \tau)$ smoothly interpolates between $0$ and $1$, at initial
and late times, i.e. $\alpha(t_o, \tau)=0$ and  $\alpha(t, \tau)=1$ (for $t\gg\tau$). 
The solution obtained with trivial ID and smooth sources (\ref{e:time_switch}) 
is referred to as $\Psi_{\mathrm{Smooth}}$.
At late times, $t\gg\tau$, a numerical function 
$\Psi^{\mathrm{N}}_{\mathrm{Jost}}$ is defined as
\bea
\Psi^{\mathrm{N}}_{\mathrm{Jost}} \equiv \Psi_{\mathrm{Impulsive}} - 
\Psi_{\mathrm{Smooth}} \,.  \label{numerical_Jost}
\eea  
The function $\Psi^{\mathrm{N}}_{\mathrm{Jost}}$ has the following 
properties~\footnote{$\salto{\chi}$ denotes a jump of a 
function $\chi$ at $r=r_p$, i.e. $\salto{\chi}\equiv \lim \limits_{\epsilon \to 0} 
\left(\chi(r_p+\epsilon)-\chi(r_p-\epsilon)\right)$.}: 
(i) It is time independent: $\partial_t\Psi^{\mathrm{N}}_{\mathrm{Jost}}=0,\forall t$.
(ii) It has a jump at the particle: $\salto{\Psi^{\mathrm{N}}_{\mathrm{Jost}}}=
- f^{-1}_{p} F(t_o,r_p)$, where $f_{p}\equiv f(r_{p})$. 
(iii) The spatial derivative, $\partial_\rstar\Psi^{\mathrm{N}}_{\mathrm{Jost}}$, is
continuous at $r=r_p$.

{\em (b) Analytical approach.}
Motivated by the numerical approach, one constructs the analytical function 
$\Psi^{\mathrm{A}}_{\mathrm{Jost}}$ as
\bea
\Psi^{\mathrm{A}}_{\mathrm{Jost}} \equiv 
\Psi^{\mathrm{A},-}_{\mathrm{Jost}}\;\Theta^{}_{-} +  
\Psi^{\mathrm{A},+}_{\mathrm{Jost}}\;\Theta^{}_{+} \,,  
\label{analytical_Jost}
\eea
where $\Theta_{+}\equiv\Theta(\rstar-\rstarp)$, $\Theta_{-}\equiv\Theta(\rstarp-\rstar)$,
being $\Theta$ the Heaviside step function, and $\Psi^{\mathrm{A},-}_{\mathrm{Jost}}$
and $\Psi^{\mathrm{A},+}_{\mathrm{Jost}}$ solve the homogeneous, stationary
version of Eq.~(\ref{e:master_equation})
\bea
\left[\partial_\rstar^2 - V(r)\right]\Psi^{\mathrm{A},\pm}_{\mathrm{Jost}} = 0 \,,
\label{e:Jost_equation}
\eea
such that $\salto{\Psi^{\mathrm{A}}_{\mathrm{Jost}}}= -f^{-1}_{p} F(t_{o},r_p)$. 
This means (see below) that $\Psi^{\mathrm{A}}_{\mathrm{Jost}}$ satisfies an 
inhomogeneous version of Eq.~(\ref{e:Jost_equation}) with a stationary
singular source given by $-f_p F(t_{o},r_p)\delta'(r-r_p)$.
Differences between $\Psi^{\mathrm{N}}_{\mathrm{Jost}}$ and 
$\Psi^{\mathrm{A}}_{\mathrm{Jost}}$ are shown to vanish within numerical 
precision in~\cite{Field:2010xn}. Thus, we refer  
to a single {\em Jost} function, $\Psi_{\mathrm{Jost}}$.

From this analysis we conclude that at late times, when the time switch-on function 
$\alpha(t, \tau)$ equals $1$ and the sources for $\Psi_{\mathrm{Impulsive}}$ and 
$\Psi_{\mathrm{Smooth}}$ coincide,  their
difference should be a solution to the homogeneous version of 
Eq.~(\ref{e:master_equation}).  However, this is in conflict with the fact that, 
as discussed above, $\Psi_{\mathrm{Jost}}$ solves a (stationary) version of 
Eq.~(\ref{e:master_equation}) with a singular distributional source. Certainly, 
this contradiction arises from the use of inconsistent sources and ID. But more 
importantly,  it also suggests strongly that, at late times, one is actually 
solving two different systems for $\Psi_{\mathrm{Impulsive}}$ and  
$\Psi_{\mathrm{Smooth}}$, namely with different sources. To assess this 
point we now discuss the implementation of Eq.~(\ref{e:master_equation}) 
in~\cite{Field:2010xn}.

In the {\em particle-without-particle} (PwP) approach to Eq.~(\ref{e:master_equation}), 
introduced in~\cite{Canizares:2008dp,Canizares:2009ay} and further discussed 
in~\cite{Canizares:2010yx,Field:2009kk,Field:2010xn}, the point mass is placed
at the boundary between two integration domains. Then, one solves a
homogeneous problem in each domain and the singular sources in 
Eq.~(\ref{e:master_equation}) are translated into {\em jump} conditions at 
the particle. To illustrate this it is convenient to rewrite Eq.~(\ref{e:master_equation}) 
as a first-order hyperbolic system (cf. Refs.~\cite{Sopuerta:2005gz,Field:2009kk} for a 
second-order formulation). To that end, we introduce the fields~\footnote{Note the 
opposite sign convention in the definition of $\Pifield$ here as compared to $\Pi$
in~\cite{Field:2009kk,Field:2010xn}.}
\bea
\label{e:First_derivatives}
\Pifield \equiv \partial_t \Psi  \,,\; 
\Phifield \equiv \partial_\rstar \Psi \,,
\eea
and Eq.~(\ref{e:master_equation}) becomes a system for the vector  
$(\Psi, \Pifield, \Phifield)$ 
\bea
\label{e:first-order-system}
\partial_t \Psi = \Pifield \,, \ \
\partial_t \Pifield  = \partial_\rstar \Phifield - V(r) \Psi - S(t,r)  \,, \ \ 
\partial_t \Phifield = \partial_\rstar \Pifield \,.
\eea
In the PwP approach we perform the splitting 
\bea
\Psi & = & \Psi^-\;\Theta^{}_{-} + \Psi^+\;\Theta^{}_{+}\,, \label{e:Heaviside_Psi} \\
\Pifield  & = & \Pifield^-\;\Theta^{}_{-}  + \Pifield^+\;\Theta^{}_{+}\,, \label{e:Heaviside_Pi} \\
\Phifield & = & \Phifield^-\;\Theta^{}_{-} + \Phifield^+\;\Theta^{}_{+} + \salto{\Psi}
\delta(\rstar-\rstarp) \,, \label{e:Heaviside_delta_Phi}
\eea
where the Dirac term of $\Phifield$ follows from consistency between the definition of $\Phifield$ 
in~(\ref{e:First_derivatives}) and the expression for $\Psi$ in~(\ref{e:Heaviside_Psi}).
Inserting expressions (\ref{e:Heaviside_Psi})-(\ref{e:Heaviside_delta_Phi}) in
system~(\ref{e:first-order-system}) we obtain homogeneous systems of equations for 
the fields
\bea
\label{e:first-order-system_homogeneous}
\partial_t \Psi^\pm  =  \Pifield^\pm \,, \ \ 
\partial_t \Pifield^\pm   =  \partial_\rstar \Phifield^\pm - V(r)\Psi^\pm  \,, \ \
\partial_t \Phifield^\pm  =  \partial_\rstar \Pifield^\pm \,, 
\eea
and jump conditions to communicate them across the particle
\bea
\salto{\Psi}(t) & = & f^{-1}_p F(t,r_p)\,, \label{e:jump_Psi}\\
\salto{\Pifield}(t)  & = & f^{-1}_p(\partial^{}_t F)(t,r_p) \,, \label{e:jump_Pi} \\
\salto{\Phifield}(t) & = & G(t,r_p) - f^{-1}_p (\partial^{}_\rstar F)(t,r_p)\,.
\label{e:jump_Phi} 
\eea
This whole system, i.e. homogeneous Eqs.~(\ref{e:first-order-system_homogeneous})
and jumps~(\ref{e:jump_Psi})-(\ref{e:jump_Phi}), is
equivalent to the original Eq.~({\ref{e:master_equation}).
Two remarks are in order, regarding the system 
(\ref{e:first-order-system_homogeneous})-(\ref{e:jump_Phi}). First,
not all initial conditions are consistent with the source $S(t,r)$,
since trivial ID violates the jump conditions~(\ref{e:jump_Psi})-(\ref{e:jump_Phi}). 
Second, jumps $\salto{\Psi}$ and $\salto{\Pifield}$ contain redundant information since, 
consistently with the definition of $\Pifield$
in (\ref{e:First_derivatives}), $\salto{\Pifield}$ is the time derivative of $\salto{\Psi}$. 
However, it is crucial to realize that the jump condition~(\ref{e:jump_Pi})
for $\salto{\Pifield}$ does not account for the initial value of $\salto{\Psi}$.
Therefore, if the evolution scheme fails to implement the condition
\be
\salto{\Psi}(t_o)= f^{-1}_p F(t_{o},r_p)\,, \label{idjumpofpsi}
\ee
it is no longer equivalent to the model Eq.~({\ref{e:master_equation}).
In~\cite{Field:2009kk,Field:2010xn}, this equation
is written in first-order form in terms of the variables $(\Psi, \Pifield, \Phifield)$,
as in (\ref{e:jump_Psi})-(\ref{e:jump_Phi}), but the equations for $\Pifield$ and
$\Phifield$ are now
\bea
\label{e:first-order-system_homogeneous_deltas}
\partial_t \Pifield  & = & \partial_\rstar \Phifield - V(r) \Psi
-J^{}_\Phifield\, \delta(\rstar -\rstarp) \,, \nn \\
\partial_t \Phifield & = & \partial_\rstar \Pifield + J^{}_\Pifield\,\delta(\rstar -\rstarp)\,,
\eea
where $J_\Phifield=\salto{\Phifield}$ and $J_\Pifield=\salto{\Pifield}$.  The jump conditions
can be recovered by multiplying the equations by a test function, integrating
(by parts) between $\rstar -\epsilon$ and $\rstar +\epsilon$, and taking the limit
$\epsilon \rightarrow 0$.
However, system (\ref{e:first-order-system_homogeneous_deltas}) does not explicitly 
enforce the jump condition on $\salto{\Psi}$. More precisely, the jump condition
on $\salto{\Psi}$ is expected to be implemented by enforcing $\salto{\Pifield}$, which 
means that $\salto{\Psi}$ is enforced up to an initial condition: Only the value 
of $\partial_t\salto{\Psi}$ is enforced.
Since $\Psi$ is coupled to the rest of the system through
the potential term, $V(r)\Psi$, this has consequences on the whole system.
In conclusion, the evolution system (\ref{e:first-order-system_homogeneous_deltas})
is not equivalent to Eq.~(\ref{e:master_equation}).

Another observation is that trivial ID is confusing: From  the source perspective 
we must impose the condition~(\ref{idjumpofpsi}), but from the 
point of view of the ID it seems reasonable to choose $\salto{\Psi}(t_o)=0$. 
This confusion is just a consequence of the inconsistency between the singular 
source term and trivial ID that we pointed out above. 

The main observation of this work is that Jost junk solutions appear as a consequence of 
implementing a {\em finite} jump condition, $\salto{\Psi}$, by enforcing an infinitesimal 
condition in time (the jump differential equation $\partial_t\salto{\Psi}=\salto{\Pifield}$), 
{\em without} simultaneously imposing the initial value of $\salto{\Psi}$ that is
consistent with the singular source [Eq.~(\ref{idjumpofpsi})].  Therefore, the
prescription to eliminate Jost junk solutions is simple: To enforce
the initial value $\salto{\Psi}(t_o)$ along the evolution so that the sources are correctly
implemented.

In light of this discussion the results and conclusions in~\cite{Field:2010xn} 
are correct.  In particular, it is concluded there that Jost junk solutions are not 
numerical artifacts but rather they are related to the implementation of the singular 
source term.  However, the discussion about the specific reason underlying the appearance 
of a Jost solution is not conclusive, as the failure to enforce the initial 
condition~(\ref{idjumpofpsi}) is not identified as the underlying cause.
Nevertheless, a way of avoiding the problem is proposed in~\cite{Field:2010xn},
consisting of a redefinition of $\Phifield$~\footnote{Note that this is precisely the 
redefinition needed for consistency  with~(\ref{e:Heaviside_delta_Phi}), in particular 
for trivial ID for $\Phifield^\pm$.}
\be
\tilde{\Phifield}= \Phifield + \salto{\Psi}\,\delta(\rstar - \rstarp) \,. 
\label{e:Phi_redefinition}
\ee
Then, the equations for $(\Pifield,\tilde\Phifield)$ become 
\bea
\label{e:first-order-system_homogeneous_deltas_2}
\partial_t \Pifield &=& \partial_\rstar \tilde{\Phifield} - V(r)\Psi
-J_\Phifield \delta(\rstar -\rstarp) -J_\Psi \delta'(\rstar -\rstarp)\,, \nn \\
\partial_t \tilde{\Phifield} &=& \partial_\rstar \Pifield  \,,
\eea
with $J_\Psi=\salto{\Psi}$. No Jost solutions are found in the implementation
of these equations with trivial ID for $(\Psi,\tilde{\Phifield},\Pifield)$, and it is concluded 
in \cite{Field:2010xn} that such kind of contamination is not {\em generic}.
The reason is clear from the conclusions we have reached:
system~(\ref{e:first-order-system_homogeneous_deltas_2}) implements the finite 
condition~(\ref{e:jump_Psi}) on $\salto{\Psi}$, including~(\ref{idjumpofpsi}), 
whereas system~(\ref{e:first-order-system_homogeneous_deltas}) does not.
Actually, system~(\ref{e:first-order-system_homogeneous_deltas_2})
is equivalent to the second-order model Eq.~(\ref{e:master_equation}), and this 
statement is independent of any discussion about ID~\footnote{Note a secondary source
of ambiguity associated with the incorrect implementation of the
{\em finite} jump $\salto{\Psi}$ that renders expression for $\salto{\Phifield}(t)$
in Eq. (13) of~\cite{Field:2009kk} (with $\dot{r}_p=\ddot{r}_p=0$)
nonequivalent to our expression~(\ref{e:jump_Phi}),
therefore introducing a potential new source of error.}.
An alternative way to understand the contrast between
systems~(\ref{e:first-order-system_homogeneous_deltas}) 
and~(\ref{e:first-order-system_homogeneous_deltas_2}) is to note  
that trivial ID for $\tilde{\Phifield}$ and $\Phifield$ have 
(cf. [17]) completely different content.  
Whereas the ambiguity stemming from the inconsistency between
singular sources and trivial ID was resolved in the case 
of~(\ref{e:first-order-system_homogeneous_deltas}) by keeping trivial ID and ignoring 
the $\delta'$ term (by ignoring the value of $\salto{\Psi}(t_{o})$), in system 
(\ref{e:first-order-system_homogeneous_deltas_2}) one instead prioritizes the
correct implementation of the singular source from $t=t_o$ at the (minimal) price 
of modifying the trivial ID on $\Phifield$ through~(\ref{e:Phi_redefinition}) and 
$\tilde{\Phifield}|_{t=t_o}=0$, which is equivalent to~(\ref{idjumpofpsi}).
In relation to this we make a statement that we discuss later:
Preserving the singular source at late times
[and this, in certain implementations of (\ref{e:master_equation}) like 
(\ref{e:first-order-system_homogeneous_deltas}) and 
(\ref{e:first-order-system_homogeneous_deltas_2}), 
depends critically on the correctness of the source at $t=t_o$] 
systematically removes the Jost junk solution, independently of the ID.

The problem discussed here is generic in the sense that it affects any linear system with 
distributional Dirac-delta sources in which (i) the sources are translated into jumps of 
the fields, and (ii) these jumps are implemented infinitesimally in time through an 
evolution equation. 
Consistency with the original system of equations demands the explicit enforcement of the 
initial value of the jump, otherwise a Jost junk solution will be present. 

To illustrate this point, we consider the computation of the self force in the simplified
scenario corresponding to a charged scalar particle orbiting a nonrotating MBH.  The
spherical harmonic modes of the retarded scalar field, $\Phi_{\ell m}$, satisfy the model 
Eq.~(\ref{e:master_equation}) with the following identifications: $\Psi=r\Phi_{\ell m}$,
$F=0$, and $G = f^{-1}S_{\ell m}$ [see, e.g.~\cite{Canizares:2009ay}, for the expressions of $V(r)$ and $S_{\ell m}(t,r)$].
With this, we can use the first-order reduction of Eqs.~(\ref{e:first-order-system}) or the one
presented in~\cite{Canizares:2010yx} based on characteristic variables, 
$(\Psi, u, v)$, with $(u,v)=(\Pifield-\Phifield,\Pifield+\Phifield)$.  Applying the splitting of the PwP
approach [Eqs.~(\ref{e:Heaviside_Psi})-(\ref{e:Heaviside_delta_Phi})] we obtain homogeneous 
equations and jump conditions:
\bea
\partial_t \Psi^\pm  =   (u^\pm + v^\pm)/2 \,,                 & & \salto{\Psi}(t) = 0\,, \nn \\
\partial_t u^\pm     =  -\partial_\rstar u^\pm - V(r) \Psi^\pm\,, & & \salto{u}(t) = -G(t,r_p)\,,\nn \\
\partial_t v^\pm     =  \partial_\rstar v^\pm - V(r) \Psi^\pm\,, & & \salto{v}(t) = G(t,r_p)\,.
\label{e:characteristic_scalar_first-order-system}
\eea
The crucial point here is the implementation of jumps $\salto{u}(t)$ and $\salto{v}(t)$.
In~\cite{Canizares:2010yx} two different strategies have been adopted: 
(I) Finite jumps  $\salto{u}(t)$ and $\salto{v}(t)$ are directly enforced in the evolution 
and trivial ID is used; 
(II) The time derivatives of the jumps, $d\salto{u}/dt$ and $d\salto{v}/dt$, are imposed
as extra evolution equations using the method of lines.  Here, trivial ID cannot be employed 
and one must {\em ``impose initially the values of the jumps [...], since during the 
evolution the only input about them is the information on their derivatives''} (cf. discussion 
in Sec.~V of~\cite{Canizares:2010yx}).  Instead, appropriately modified ID is used
[note the similarity to the discussion of system~(\ref{e:first-order-system_homogeneous_deltas_2})].

Therefore, a natural question is whether or not a Jost junk solution appears when 
taking the difference between the solutions obtained by implementing
approaches (I) and (II) but using trivial ID in {\em both} cases.
Figure~\ref{f:Figure1} shows the results obtained from the numerical implementation
with the techniques discussed in Ref. \cite{Canizares:2010yx},
which confirm our conclusion that Jost junk solutions correspond to incorrect 
implementations of the distributional sources, rather than to (trivial) ID inconsistent 
with the sources.
\begin{figure}[t]
\centerline{\includegraphics[scale=0.7]{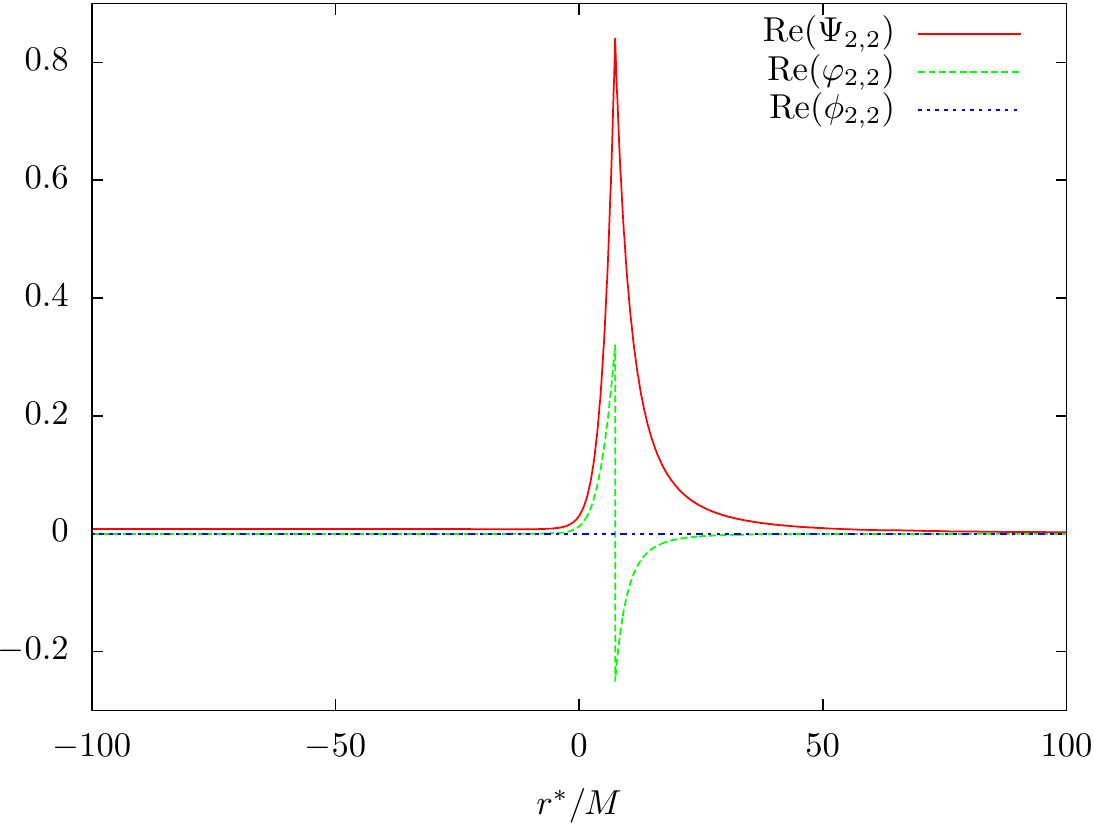}}
\caption[]{\label{f:Figure1}  
{\em Jost} solution built  by subtracting 
  solutions to system (20) in approaches (I) and (II) for jump 
  implementation (cf. text), and using trivial ID in {\em both} cases.}
\end{figure}

Hitherto, we have focused mainly on consistency issues {\em at initial times}. 
In order to get a deeper insight, 
we paraphrase the main discussion above in terms of solutions {\em at late times}.
The only relevant element in the analysis at late times
is the correct implementation of the sources, the choice of ID playing no role. 
In particular, any intermediate-time data can be taken as valid ID.
From this perspective, the crucial impact of the failure in passing the initial 
value of the field jumps~\footnote{The initial 
value of the jump is part of the implementation of the sources and hence, 
it is prior to and independent from the choice of ID.
Only in some particular implementations, like the one in~\cite{Field:2010xn}, 
both issues are linked.} is the incorrect 
implementation of the sources, which differ from the correct ones at late times. 
Then, one is solving a different problem. 

The reason why a Jost solution appears when evaluating $\Psi^{N}_{\mathrm{Jost}}$ 
[Eq.~(\ref{numerical_Jost}})] in~\cite{Field:2010xn} is because the smooth time switch-on
function,
$\alpha(t,\tau)$, guarantees the correct implementation of the late time sources
in the solution $\Psi_{\mathrm{Smooth}}$ when adopting a strategy in which the source 
is implemented through an evolution
equation for the jumps with zero initial values. 
To illustrate this, let us consider a field $\chi$ with jump condition
$\salto{\chi}=J_{\chi}(t)$ and its smoothed version $\salto{\tilde\chi} = 
\alpha(t,\tau)J_{\chi}(t)$.  If we implement the jumps using evolution equations,
$d\salto{\chi}/dt = J'_{\chi}$ and $d\salto{\tilde\chi}/dt= (\alpha J_{\chi})'$
respectively, both with zero initial values, $\salto{\chi}(t_o)=\salto{\tilde{\chi}}(t_o)=0$,
we obtain
\bea
\salto{\chi} & = &  J_{\chi}(t) - J_{\chi}(t_o) \,, \nn \\
\salto{\tilde{\chi}} & = &  \alpha(t,\tau)J_{\chi}(t) - \alpha(t_o,\tau)J_{\chi}(t_o)\nn \\
& = & \alpha(t,\tau)J_{\chi}(t) \simeq J_{\chi}(t)~(\mbox{for}~t\gg\tau)\,. 
\label{e:sources_late_time}
\eea
This shows conclusively that $\Psi_{\mathrm{Smooth}}$ solves Eq.~(\ref{e:master_equation})
at late times with the correct source term, whereas $\Psi_{\mathrm{Impulsive}}$ is contaminated by a Jost
solution [cf. point (ii) in the discussion after Eq. (\ref{numerical_Jost})].

{\em Time switch on in self-force calculations.} 
Although the smooth switch on is not the critical element in
the discussion of the Jost solution
\footnote{The critical element is the correct 
implementation of the source term at late times, the smooth switch on just providing a
particular manner of guaranteeing this.},
it certainly improves the consistency between the
source and the ID [which can be critical in certain implementations of the
model Eq.~(\ref{e:master_equation})].  First, it reduces the initial burst
of junk radiation, addressed in detail in~\cite{Field:2010xn}.  Second, it 
improves the calculation of the self force in certain implementations.  
In~\cite{Field:2010xn} a similar point
was addressed: The calculation of the gauge invariant gravitational-wave fluxes of energy and 
angular momentum.  Regarding the self force itself, we have computed it for the case
of a charged scalar particle in circular orbits using the {\em mode sum} regularization 
scheme~\cite{Barack:2001gx}.  The retarded field is evolved using the 
system~(\ref{e:characteristic_scalar_first-order-system}) and trivial ID in the approach
 (I), i.e. with the
direct enforcement of the {\em finite} jumps $\salto{u}$ and $\salto{v}$
(this is an example of a correct implementation of the source term while
using trivial ID and, as expected, no Jost solution appears).  On the other hand, 
the instantaneous switch on of the source term produces high-frequency numerical
noise with a very slow time decay (the decay time scale is
much larger than the orbital time scale) 
and deteriorates significantly the accuracy of the self force.
This noise can be eliminated by increasing the resolution of the computations or
by using a numerical filter.  However, a more efficient and better adapted method is the smooth
switch on of the source, which produces dramatic improvements in the computation
of the self force. In Fig.~\ref{f:Figure2} we illustrate the calculation 
of the self force, with and without time switch-on of the sources, for a
particle at the last stable circular orbit ($r_p=6M$) 
after two orbits, i.e. for $t>2\cdot T_{\mathrm{orb}}=2\cdot 
2\pi ({r_p^3/M})^{\frac{1}{2}}\approx 184.69 M$
(the self force calculation is physically meaningful at late times, 
once the initial unphysical burst has been radiated away from the particle).
Outgoing boundary conditions discussed in 
\cite{Canizares:2009ay,Canizares:2010yx} are imposed at the boundaries 
$[\rstar_{\mathrm{inner}} \approx -292 M, \rstar_{\mathrm{outer}} \approx 307 M]$,
whose location guarantees their causal disconnection from the particle
up to the final evolution time $t_{\mathrm{fin}} = 250M$.
\begin{figure}[t]
\centerline{\includegraphics[scale=0.7]{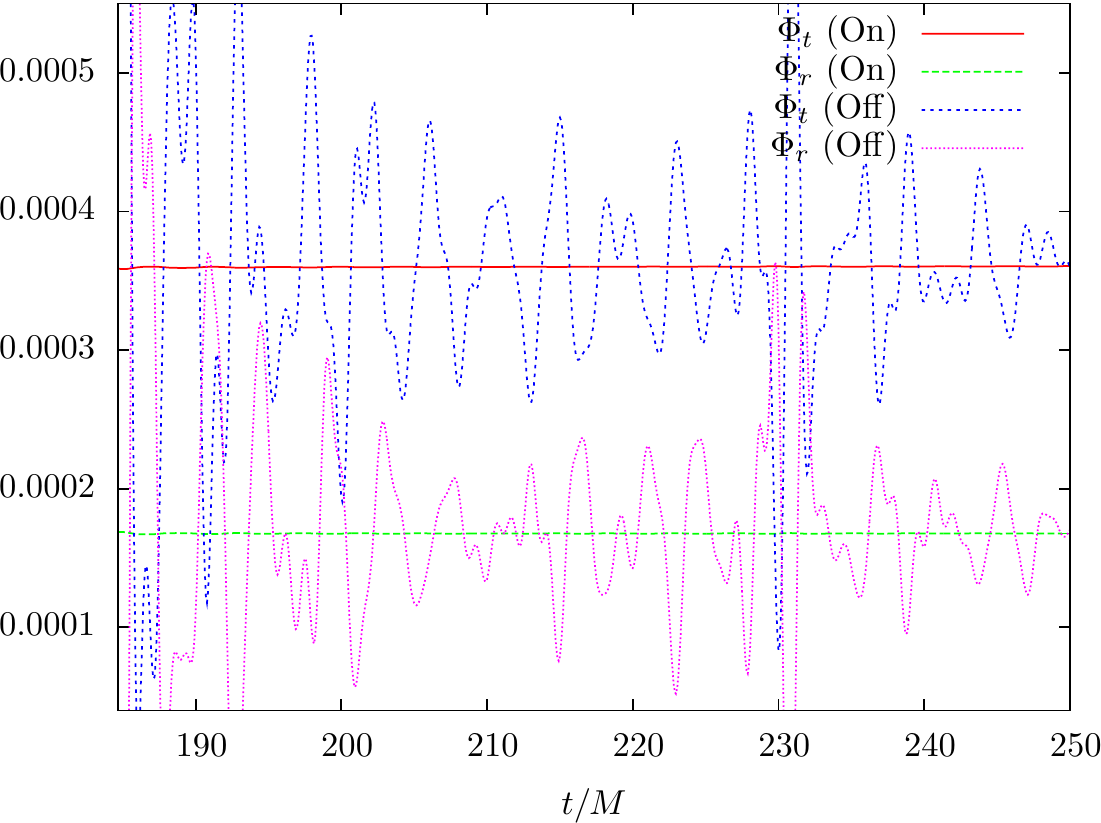}}
\caption[]{\label{f:Figure2}  
Radial and time components of the self force as a function of time,
with (``On'') and without (``Off'') time switch-on of the sources ($r_p=6M$).
Note that the Off case does not even allow
the recovery of the second significant digit in $\Phi_r=1.677 \cdot 10^{-4} $, 
$\Phi_t=3.609 \cdot 10^{-4}$ (see e.g.~\cite{Canizares:2009ay}).}
\end{figure}
%

To conclude, we have analyzed the risk of contamination of time-domain self-force calculations 
by Jost junk solutions as a consequence of using trivial ID inconsistent with the
distributional sources, which has produced some confusion in the community.
We have shown that no Jost solution is generated as long as 
the distributional Dirac-delta sources are faithfully implemented at late times.
Indeed, the inconsistency between trivial ID and distributional sources introduces
a genuine ambiguity in the evolution system: either one modifies the sources to make them
compatible with the trivial ID, or one accepts the use of inconsistent ID and prioritizes
the implementation of the correct sources (at late times). 
In this specific sense, the conclusion in~\cite{Field:2010xn} stating that the presence 
of Jost solutions is not a numerical artifact, but actually depends on the implementation
of the system, is correct. However, it is not a generic feature of these systems and hence
can be easily avoided.  Here, we have first identified unambiguously the specific origin of 
Jost solutions in implementations of jump conditions due to Dirac-delta sources, namely,
the failure to enforce the correct initial value in the evolution equations for the jumps.
Second, our final conclusion is that no contamination of the retarded solution
by Jost junk solutions happens as long as (late time) sources are correctly
implemented, and this is conceptually independent of the ID and/or the use 
of a smooth time switch-on.  
As a by-product of the analysis, we have shown that the use of a smooth time switch-on of 
the sources~\cite{Field:2009kk,Field:2010xn} prevents
the high-frequency noise that spoils self-force calculations in some numerical schemes.

We would like to thank A. Harte and B. Wardell for insightful comments. 
J.L.J. acknowledges support from the Alexander von Humboldt Foundation. 
C.F.S. acknowledges support from the Ram\'on y Cajal Programme of the Spanish Ministry 
of Education and Science (MEC) and by a Marie Curie International Reintegration Grant 
No. MIRG-CT-2007-205005/PHY (FP7). 
P.C.M. is supported by the Spanish Ministry of Science and Innovation (MICINN).
We also acknowledge financial support from Contracts No. ESP2007-61712 (MEC),
No. FIS2008-06078-C03-01/03 (MICINN), and No. FQM2288 and No. FQM219 (Junta de Andaluc\'{\i}a).


\end{document}